\documentclass[aps,prl,showpacs,twocolumn,floatfix,nofootinbib,superscriptaddress]{revtex4-1} 

\usepackage{amsmath,amssymb,amsfonts,graphicx}
\usepackage{bm}
\usepackage{slashed}
\usepackage{subfigure}
\usepackage[svgnames]{xcolor}

\newcommand{\vect}[1]{\boldsymbol{#1}}

\begin{document}

\title{Higher Order Collins Modulations in Transversely Polarized Quark Fragmentation}

\preprint{ADP-12-22/T789}
%{\raggedright ADP-12-22/T789}

\author{Hrayr~H.~Matevosyan}
\affiliation{ARC Centre of Excellence for Particle Physics at the Tera-scale,\\ 
and CSSM, School of Chemistry and Physics, \\
The University of Adelaide, Adelaide SA 5005, Australia
%\\ http://www.physics.adelaide.edu.au/cssm
}

\author{Anthony~W.~Thomas}
\affiliation{ARC Centre of Excellence for Particle Physics at the Tera-scale,\\     
and CSSM, School of Chemistry and Physics, \\
The University of Adelaide, Adelaide SA 5005, Australia
%\\ http://www.physics.adelaide.edu.au/cssm
}

\author{Wolfgang Bentz}
\affiliation{Department of Physics, School of Science,\\  
Tokai University, Hiratsuka-shi, Kanagawa 259-1292, Japan
%\\ http://www.sp.u-tokai.ac.jp/
}

\begin{abstract}
The Collins effect describes the modulation of the hadron production  
by a transversely polarized quark with the sine of the polar angle, $\varphi$,
between the produced hadron's transverse momentum and the quark spin. 
We employ a quark-jet model to describe multiple hadron emissions by 
such a quark, taking the Collins effect into account. The resulting hadron 
distributions exhibit modulation up to fourth order in $\sin(\varphi)$ when 
only two hadron emissions are allowed, rising with any further increase in the 
number of emitted hadrons. These new effects are a direct consequence of the 
quark-jet mechanism for quark hadronization, which do not depend on the details 
of the model used for elementary hadron emission. The size and the sign of 
the higher order terms are directly connected with the probabilities of 
quark spin flip in the elementary emission process, with opposing sign favored
and unfavored Collins functions only being generated if quark spin flip is preferential.
Experimental studies of these effects should therefore provide a critical 
test of the quark hadronization mechanism, which in turn will lead to a 
deeper understanding of the transverse spin structure of hadrons.

\end{abstract}


\date{\today}                                           % Activate to display a given date or no date

\maketitle
%%%%%%%%%%%%%%%%%%%%%%%%%%%%%%%%%%%%%%%%%%%%%%%%%%%%%
%%%%%%%%%%%%%%%%%%%%%%%%%%%%%%%%%%%%%%%%%%%%%%%%%%%%%
%%%%%%%%%%%%%%%%%%%%%   SECTION %%%%%%%%%%%%%%%%%%%%%%%%%%
%\section{Introduction}
%
Quark hadronization remains one of the most challenging problems in hadronic 
physics, as it involves both long and short-range parton interactions. 
This process is quantitatively described by various fragmentation functions, 
which have a probabilistic interpretation. For example, the unpolarized 
fragmentation function, $D_1^{h/q}(z)$, can be interpreted as the probability 
density for an unpolarized quark $q$ to emit a hadron $h$ carrying 
light-cone momentum fraction $z$. These functions enter the expressions for 
the cross-sections of various hadronic processes in the factorization regime, 
such as semi-inclusive deep inelastic scattering (SIDIS) and $e^+ e^-$ and 
hadron-hadron collisions, convoluted where necessary with the relevant 
parton distribution functions. Thus, a reliable knowledge of the fragmentation 
functions is crucial for disentangling the underlying parton distribution functions.

In recent years there has been particular interest in studying scattering 
processes with transversely polarized probes and targets. Here the fragmentation 
of the transversely polarized quarks to spin zero particles, like pions 
and kaons, can be described by two functions: the familiar unpolarized 
fragmentation function and the na\"ively time-reversal-odd Collins 
function~\cite{Collins:1985ue,Collins:1989gx}. 

\begin{figure}[b]
\centering\includegraphics[width=0.44\textwidth]{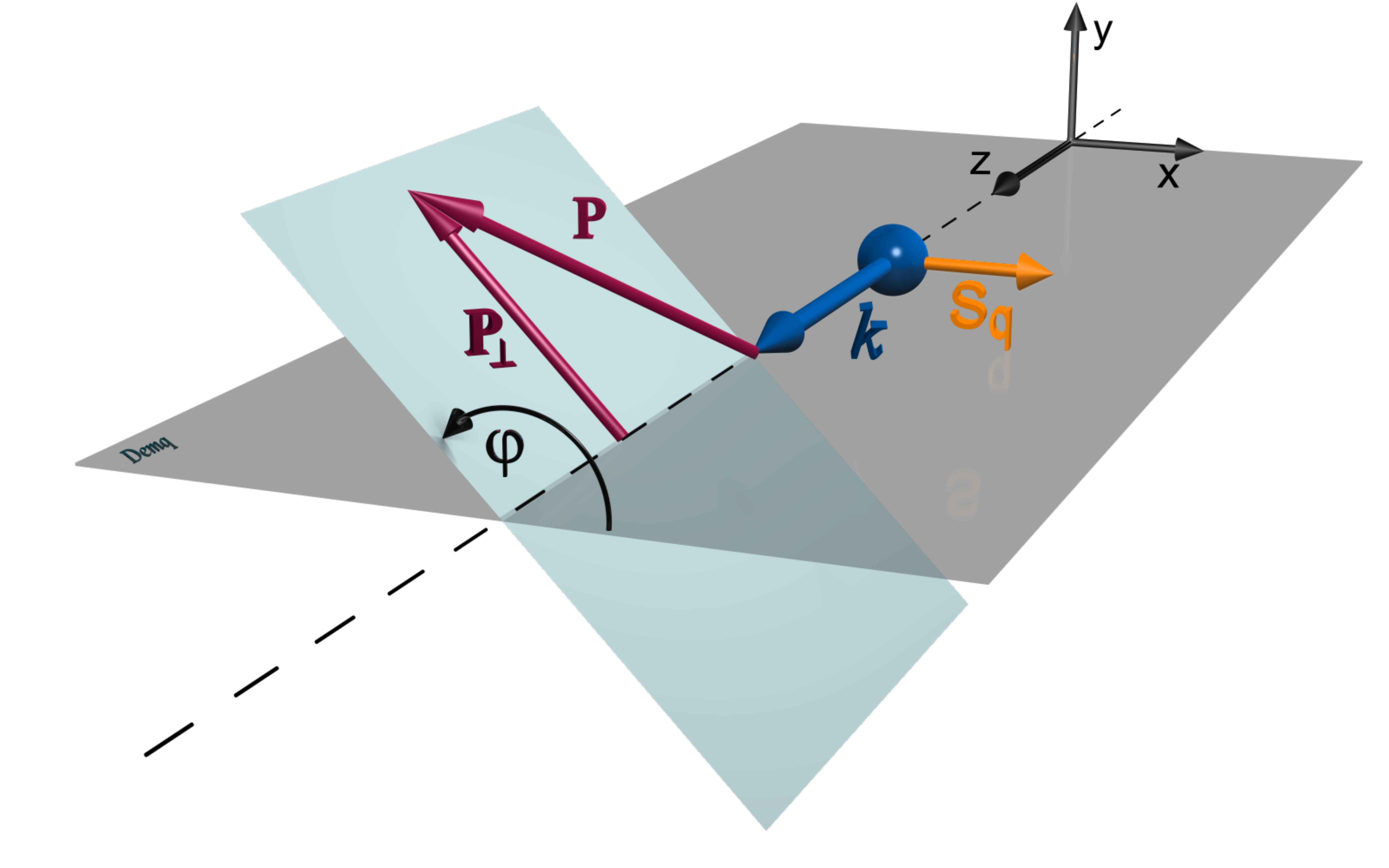}
\caption{Illustration of transversely polarized 
quark fragmentation. 
The quark momentum, $\vect{k}$, defines the $z$-axis, 
with its transverse polarization spin vector, $\vect{S}_q$, 
along the $x$ axis. 
The emitted hadron has momentum $\vect{P}$ with component $\vect{P_{\perp}}$ 
transverse with respect to the $z$-axis. The polar angle of  
$\vect{P_{\perp}}$ with respect to $\vect{S}_q$ is denoted $\varphi$.}
\label{PLOT_POL_QUARK_3D}
\end{figure}

The probability of the transversely polarized quark $q$ to emit a spin-zero 
hadron $h$ with transverse momentum $\vec{P}_\perp$,  depicted schematically 
in Fig.~\ref{PLOT_POL_QUARK_3D}, can be expressed as~\cite{Bacchetta:2004jz}
\begin{align}
\label{EQ_Dqh_SIN}
D_{h/q^{\uparrow}} (z,P_\perp^2,\varphi) &= D_1^{h/q}(z,P_\perp^2)\\ \nonumber
 &- H_1^{\perp h/q}(z, P_\perp^2) \frac{ P_\perp S_q}{z m_h} \sin(\varphi) \, ,
\end{align}
where $m_h$ is the mass of the produced hadron.
The transverse momentum dependent (TMD) unpolarized fragmentation function
is denoted by $D_1^{h/q}(z,P_\perp^2)$, while $H_1^{\perp h/q}(z, P_\perp^2)$ 
is the Collins function. Thus the Collins function modulates the distribution of
produced hadrons with $\sin(\varphi)$, a clear signature that allows one 
to extract it from the experimental data. The experimentally measured Collins 
functions from HERMES, COMPASS and JLab strongly suggest that the 
$1/2$ moments
\begin{equation}
H_{1 (h/q)}^{\perp  (1/2)}(z) \equiv 
\pi \int_0^{\infty} d P_\perp^2 \frac{P_\perp}{2z m_h}  
H_1^{\perp h/q}(z, P_\perp^2) \,
\label{eq:halfmoment}
\end{equation}
of the unfavored Collins functions have a similar size and opposite sign to those for the favored 
ones for pions~\cite{Avakian:2003pk,Airapetian:2004tw,Bradamante:2011xu,
Aghasyan:2011ha,Aghasyan:2011gc}. 
We note that the fragmentation functions are called favored, where the produced 
hadron has a valence quark of the same flavor as the initial fragmenting quark, 
and  unfavored otherwise. There has been extensive modeling of the Collins 
function in the 
past~\cite{Amrath:2005gv,Bacchetta:2007wc, Gamberg:2003eg, Artru:2010st}. 
Most notably, the spectator model calculations of 
Refs.~\cite{Amrath:2005gv,Bacchetta:2007wc, Gamberg:2003eg} provide a microscopic 
description of the fragmentation process. However, these calculations have been 
limited to the single hadron emission approximation, with the consequence 
that the unfavored fragmentation functions are zero in this approach. In 
addition, the description of the small $z$ region for the favored fragmentation 
functions requires the introduction of a phenomenological quark-meson form factor.
 
\begin{figure}[tbp]
\centering\includegraphics[width=0.44\textwidth]{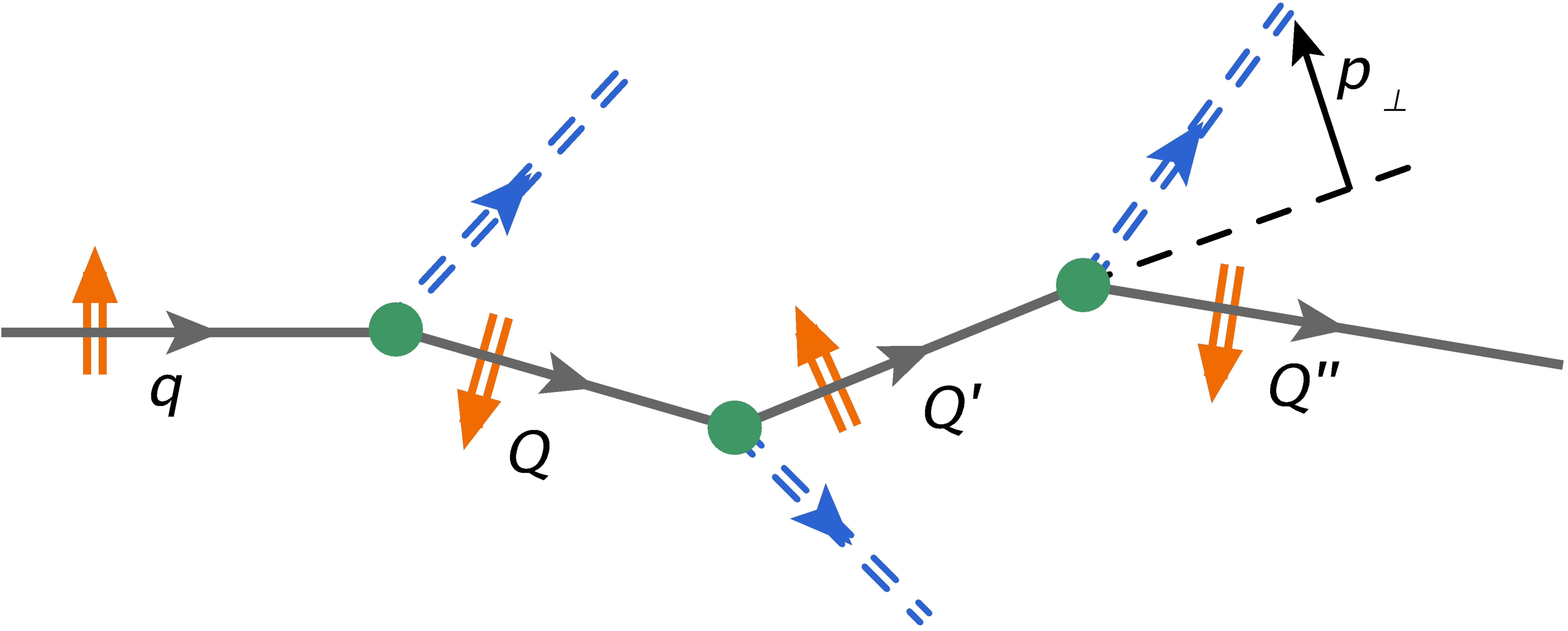}
\caption{NJL-jet model including transverse momentum and quark polarization 
transfer. Here the (orange) double-lined arrows indicate the quark's spin.}
\label{PLOT_NJL-JET_TMD}
\end{figure}

Recently, we employed the NJL-jet model of 
Refs.~\cite{Ito:2009zc,Matevosyan:2010hh,Matevosyan:2011ey,Matevosyan:2011vj} 
to calculate the Collins fragmentation function within the quark-jet 
hadronization picture~\cite{Matevosyan:2012ga}, using Monte Carlo (MC) 
simulations. In this picture, the initial quark undergoes a decay chain of 
hadron emissions, schematically depicted in Fig.~\ref{PLOT_NJL-JET_TMD}, 
where a microscopic description of each elementary emission is provided by
the effective quark model of Nambu and Jona-Lasinio (NJL)~\cite{Nambu:1961tp}.  
  
In this Letter we show that within this picture  
the fragmentation function of a transversely polarized quark acquires 
Collins modulations of higher order in $\sin(\varphi)$ as a direct consequence 
of the quark-jet hadronization process.
These new terms are essential to describe the unfavored fragmentation functions. 
It appears that, even with just two hadron emissions in the decay chain, 
one requires a fourth order polynomial in $\sin(\varphi)$ to fully describe the 
produced number densities.
We stress that this effect is independent of the particular form of the model 
used to describe the Collins effect at the elementary emission vertex 
but is a direct consequence of the multiple hadron emission mechanism.

%===============================================================================
%===============================================================================
  
%%%%%%%%%%%%%%%%%%%%%%%%%%%%%%%%%%%%%%%%%%%%%%%%%%%%%
%%%%%%%%%%%%%%%%%%%%%%%%%%%%%%%%%%%%%%%%%%%%%%%%%%%%%
%%%%%%%%%%%%%%%%%%%%%   SECTION %%%%%%%%%%%%%%%%%%%%%%%%%%
%\section{Higher Order Collins Modulations in Quark-jet Model.}
%\label{SEC_MODULATIONS}
% 
In the NJL-jet model of Matevosyan {\it et al.}~\cite{Matevosyan:2012ga} 
we used the tree level diagrams to calculate the unpolarized fragmentation 
functions (the first term in Eq.~(\ref{EQ_Dqh_SIN})), while for the 
Collins functions we used the spectator model 
of Bacchetta {\it et al.}~\cite{Bacchetta:2007wc}. In that work, the interference 
term between a tree level amplitude and the amplitude with gauge link coupling 
was used to produce non-zero Collins function.
%===============================================================================
%=============================================================================== 

A key element of this calculation was the evaluation of the quark spin-flip 
probability in each hadron emission step. We used light-cone spinors to describe 
the initial and final polarized quark wave-functions, and found that the 
quark spin non-flip and spin flip probabilities are proportional to
\begin{align}
\label{EQ_SPIN_FLIP-NON}
| a_1 |^2\sim {l_x^2},\ | a_2 |^2 \sim {l_y^2+(M_2-(1-z)M_1)^2},
\end{align}
where $M_{1(2)}$ is the constituent mass of the fragmenting (remnant) quark  
and $\vect{l}_\perp$ is the transverse momentum of the remnant quark.
(Note that this momentum cancels the transverse momentum of the emitted hadron 
with respect to the momentum of the fragmenting quark.) Here we clearly 
see that the quark spin-flip probability should be on average larger than the 
spin non-flip probability. This will be shown later to be crucial  
in describing the Collins function.

\begin{figure}[phtb]
\centering 
\includegraphics[width=0.44\textwidth]{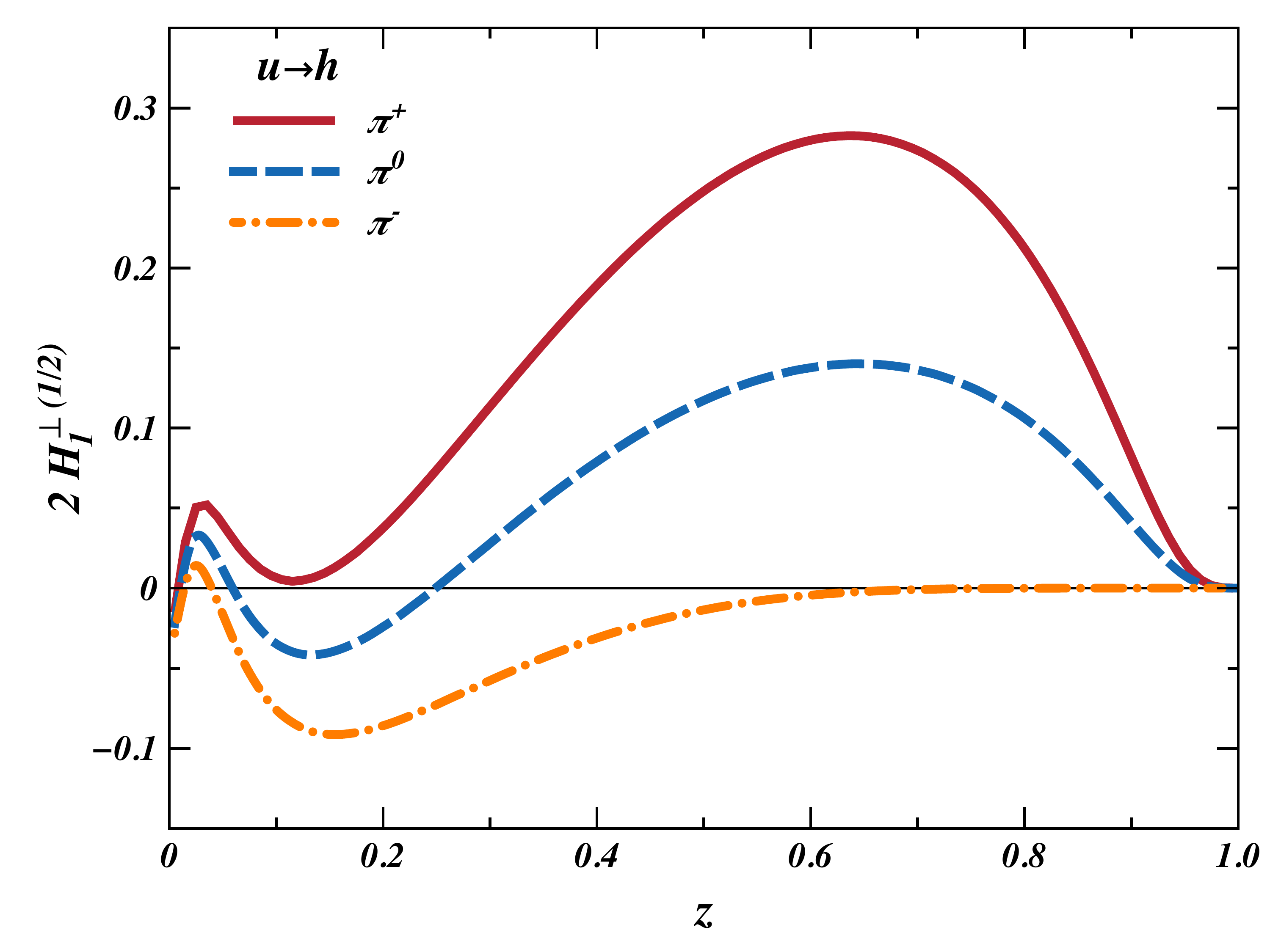}
\caption{Fitted values for $2H_1^{\perp (1/2)}$ for pions produced by a $u$ quark from MC simulations with $6$ emitted hadrons.}
\label{PLOT_H12_U}
\end{figure}

We use the elementary fragmentation function for a transversely 
polarized quark calculated within the NJL model, 
along with the quark spin flip probabilities explained above, as the 
input to a Monte Carlo simulation of the quark hadronization process 
within the quark-jet model -- see  
 Ref.~\cite{Matevosyan:2012ga} for a more detailed discussion. 
We used the $z$, $P_\perp^2$ and $\varphi$ dependence of the resulting hadron 
number densities to extract the unpolarized and Collins functions for pions and 
kaons produced by light and strange quarks. The transverse momentum integrated 
number densities were perfectly described by a first order polynomial 
in $\sin(\varphi)$ and hence the constant term was interpreted as the 
integrated unpolarized fragmentation function, while the coefficient of the 
linear term as twice the $1/2$ moment of the Collins function 
(c.f. Eq.~(\ref{eq:halfmoment})). As perhaps the most remarkable result, we
note that the unfavored Collins function was found to have the opposite sign 
and a magnitude comparable to the favored one for pions, as can be seen
from the plot in Fig.~\ref{PLOT_H12_U}.

 %===============================================================================
%===============================================================================

\begin{figure}[tbp]
\centering\includegraphics[width=0.44\textwidth]{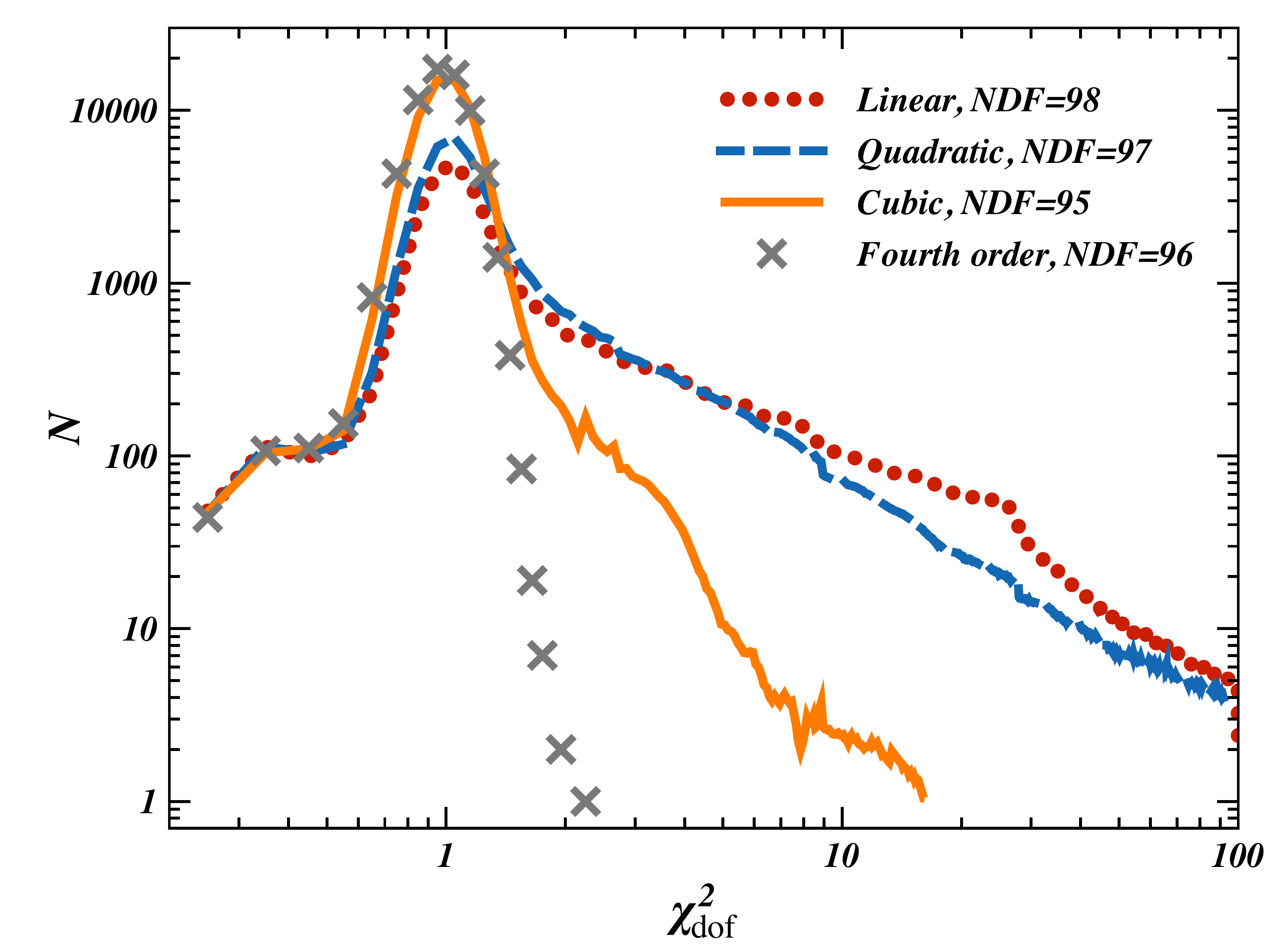}
\caption{Histogram of the values of $\chi^2_{\rm dof}$ for all fragmentation functions
of $u$  quark fitted with polynomials of different order 
in $\sin(\varphi)$ for MC simulations with $2$ emitted hadrons.}
\label{PLOT_CHI-SQ_MODEL}
\end{figure}

Even more interesting effects were observed when we considered the extraction 
of the TMD Collins function from the unintegrated number densities. 
We used a polynomial form in $\sin(\varphi)$ to minimise $\chi^2$  for a 
fit to the $\varphi$ dependence for fixed values of $z$ and $P_\perp^2$. 
Here, even for only two hadron emissions in the jet, the fits with just 
a linear form were insufficient. On the other hand,
a fourth order polynomial in $\sin(\varphi)$ provides an excellent fit everywhere. 
The plots in Fig.~\ref{PLOT_CHI-SQ_MODEL} illustrate the $\chi^2$ per degree 
of freedom ($\chi^2_{\rm dof}$) for fits with polynomials of different orders for 
all the fragmentation functions of $u$ quark (to seven pions and kaons in total)
 and for all the discrete values of $z$ and $P_\perp^2$ in our simulations -- a total of
$7\times100\times100=7\cdot 10^4$ data points. Clearly the fourth order 
polynomial yields $\chi^2_{\rm dof}$ in the vicinity of $1$ for all the fits.  
A similar analysis for the $P_\perp^2$ integrated number densities shows that 
a simple first order polynomial describes all the data perfectly well, meaning 
that this effect is only present in the TMD fragmentations.
%===============================================================================
%=============================================================================== 
 
These higher order modulations can be attributed to two effects. First, in the 
process of multiple hadron emissions, the distribution of the $n$-th produced 
hadron is multiplied  by an additional factor of $(a+b\sin(\varphi))^{(n-1)}$, 
when averaged over all Monte Carlo simulations. This can be easily deduced from 
Eq.~(\ref{EQ_Dqh_SIN}), as with each hadron emission the remnant quark number 
densities, relative to the fragmenting quark ($d_q^Q$), are related to the number
density 
of the emitted hadron ($d_q^h$) by the momentum and flavor conservation law
$d_q^Q(1-z,-\vect{p}_\perp)\propto d_q^h(z,\vect{p}_\perp)$, where $\vect{p}_\perp$
is the transverse momentum of the produced hadron with respect to the fragmenting
quark. Thus these quark densities also get modulated with a first 
order polynomial form. The next emitted hadron's distribution is modulated 
not only by the elementary Collins function, but also by the number density of 
the fragmenting quark, yielding a quadratic $\sin(\varphi)$ term in the angular 
dependence of not only the produced hadron but also the next remnant quark in the 
decay chain. Consequently, from this effect alone the distribution of each 
consecutive emitted hadron acquires another order in $\sin(\varphi)$. 
For a detailed investigation of this effect we refer to 
Matevosyan {\it et al.}~\cite{Matevosyan:2012spin},
who employed a toy model to study the effect of quark spin flip in the decay 
chain for unfavored fragmentation functions. 

The second source of the angular modulation arises from the spin flip probabilities 
themselves. This can be easily seen from the expressions for the spin flip 
probability of Eq.~(\ref{EQ_SPIN_FLIP-NON}), where $\vec{l}_\perp = -\vec{p}_\perp$ and $p_{x}= p_\perp \cos(\varphi)$, $p_{y}= p_\perp \sin(\varphi)$.
Thus both $| a_1 |^2$ and $| a_2 |^2$ are modulated with a polynomial of the 
form $a+b \sin^2(\varphi)$. Together these effects create an angular modulation 
of the hadrons produced at $n$'th order that is a polynomial in $\sin(\varphi)$ 
of order $1+3\times (n-1)$. Then we can write the general form for the polarized
fragmentation function for $R$ hadron emissions as 
\begin{align}
\label{EQ_COL_GEN}
D_{h/q^{\uparrow}} (z,P_\perp^2,\varphi) = \sum_{n=0}^{1+3(R-1)} c_n(z, P_\perp^2) \sin^n \varphi.
\end{align}

Thus, in the case with only two hadron emissions, 
we need a fourth order polynomial to fully describe the angular modulation, as was shown in our $\chi^2$ fits in Fig.~\ref{PLOT_CHI-SQ_MODEL}.

\begin{figure}[h]
\centering 
\subfigure[] {
\includegraphics[width=0.44\textwidth]{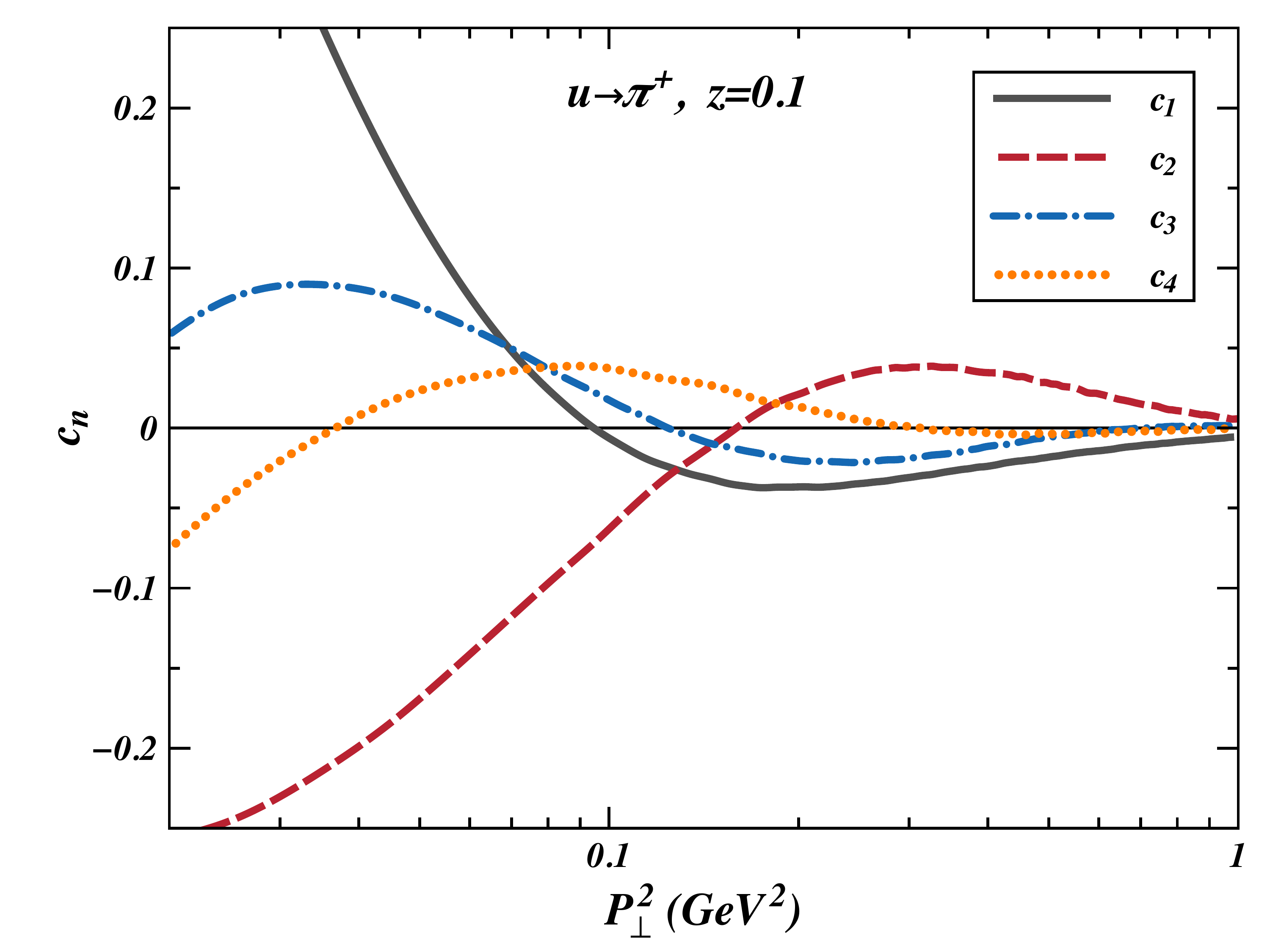}
}
\vspace{-0.3cm} 
\subfigure[] {
\includegraphics[width=0.44\textwidth]{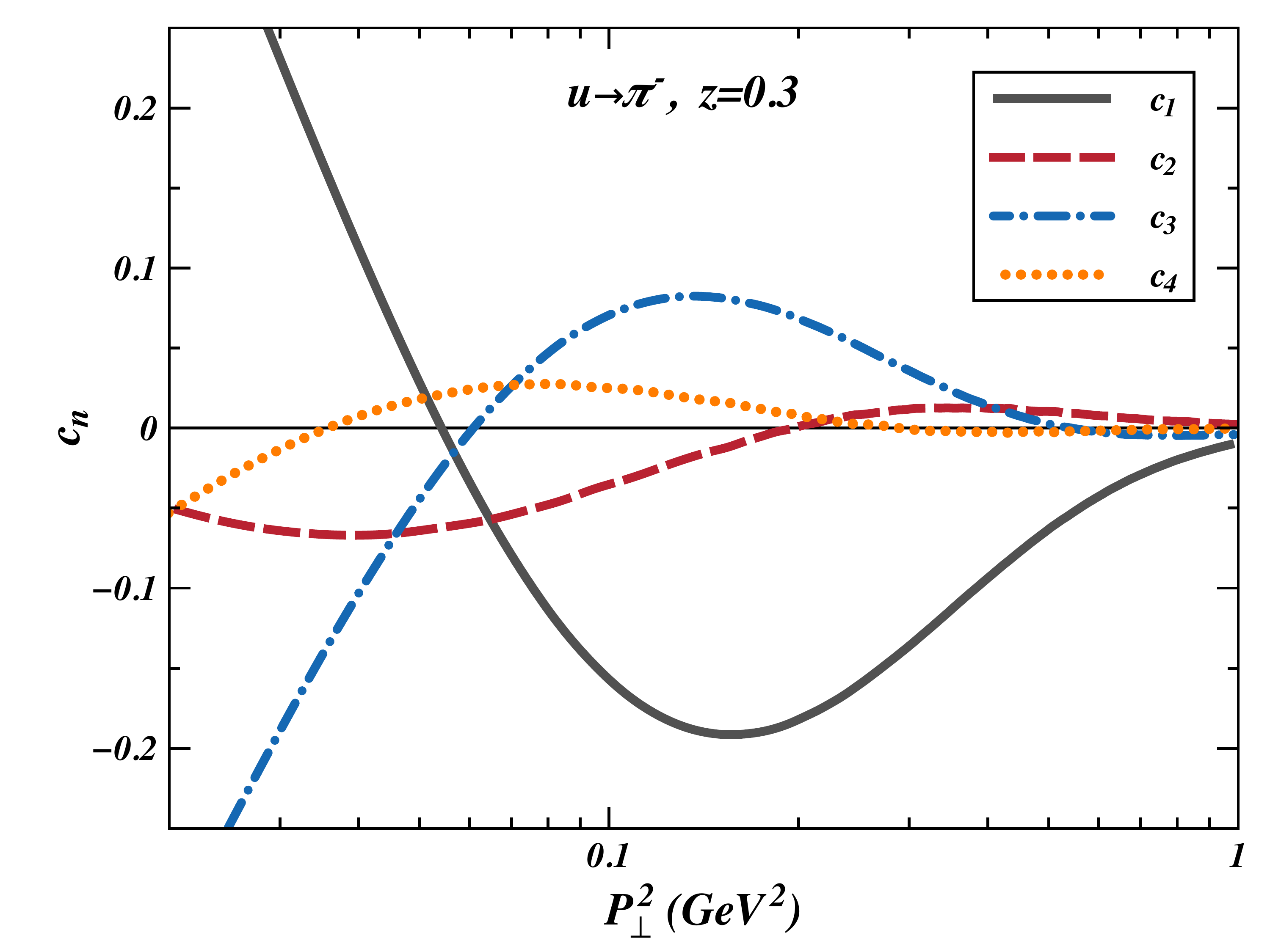}
}
\caption{The fitted values for the polynomial coefficients in $\sin(\varphi)$, 
$c_n$, (a) for $\pi^+$ at $z=0.1$  
and (b) for $\pi^-$ at $z=0.3$,   
produced by a $u$ quark for simulations with $6$ emitted hadrons.}
\label{PLOT_U_PI}
\end{figure}

Figure~\ref{PLOT_U_PI} shows the coefficients, $c_n$, of the $n$'th power 
of $\sin(\varphi)$, obtained from a $\chi^2$ fit to the fragmentation function 
for a $u$ quark going to a $\pi^+$ at $z=0.1$ and to a $\pi^-$ at $z=0.3$, 
as a function of $P_\perp^2$. In this case the Monte Carlo simulations involved
six emitted hadrons (beyond which the results are stable). Here we do not show 
the constant term corresponding to the unpolarized fragmentation function, as it 
is much larger and has been extensively studied in our previous 
work~\cite{Matevosyan:2011vj, Matevosyan:2012ga}. 
The linear term, $c_1$, corresponds to the Collins function. The plots clearly 
show that the higher order terms follow the pattern set by the linear Collins 
term studied in Ref.~\cite{Matevosyan:2012ga}: they  vanish for large values 
of $P_\perp^2$ and oscillate with an increasing magnitude with $P_\perp^2\to0$. 
In addition, the higher order terms are comparable in magnitude to the 
Collins terms in the small $z$ region. It is worth noting, that the 
higher order terms for unfavored fragmentation remain significant compared 
to the regular Collins term to substantially larger values of $z$ than 
the favored fragmentations.
%===============================================================================
%===============================================================================

We note that the hadrons produced further down the quark decay 
chain carry increasingly smaller momentum fractions and thus impact the 
distribution functions only in the low $z$ region.
This was shown explicitly for the integrated, unpolarized fragmentation function 
in Ref.~\cite{Matevosyan:2011ey}, where the plots in Fig.~5 demonstrate that 
the fragmentation functions hardly change after just three hadron emissions 
in the region $z \gtrsim0.2$. Thus the higher order modulations of the 
distributions appear mainly in the low $z$ region, making it harder to 
isolate the higher order polynomial terms in the MC data. Thus, a fourth order
polynomial was sufficient to describe our MC results within the statistical errors 
with $6$ hadron emissions.
%%%%%%%%%%%%%%%%%%%%%%%%%%%%%%%%%%%%%%%%%%%%%%%%%%%%%
%%%%%%%%%%%%%%%%%%%%%%%%%%%%%%%%%%%%%%%%%%%%%%%%%%%%%
%%%%%%%%%%%%%%%%%%%%%   SECTION %%%%%%%%%%%%%%%%%%%%%%%%%%
%\section{Conclusions}
%\label{SEC_CONCLUSIONS}
% 

In this work we have described for the first time the higher order Collins 
modulation effects for the hadrons produced in the fragmentation of a 
transversely polarized quark within the quark-jet hadronization picture. 
We showed that these higher order polynomial terms in $\sin(\varphi)$ necessarily 
arise from two effects in each hadron emission step: the linear Collins modulation 
of the remnant quark number densities in the decay chain and the $\sin^2(\varphi)$ 
modulation of the quark spin flip probabilities. In principle, this leads  
to an increase of the angular modulations of the resulting hadron distributions
by three orders in $\sin(\varphi)$ with each hadron emission after the first one. 
We showed that with just a linear Collins modulation as an input in the elementary 
quark emission our Monte Carlo results require a fit with a polynomial of 
at least fourth order to achieve satisfactory values of $\chi^2_{\rm dof}$. 
Earlier toy model tests in Ref.~\cite{Matevosyan:2012spin} helped to establish 
this mechanism of generating higher order modulations. Moreover, they showed that 
the experimental suggestion that the $1/2$ moments of the favored and unfavored 
Collins functions may have opposite signs, may be understood in the quark-jet 
picture if, on average, in the elementary hadron emission step the remnant quark's
spin is antiparallel to that of the fragmenting quark. 
  
These new Collins modulation effects are only visible in the transverse momentum 
dependent hadron number densities and grow relative to the unpolarized and 
regular Collins terms in the small $z$ and $P_\perp^2$ region, where the multiple 
hadron emission effects are the largest. Nevertheless, if observed experimentally, 
these effects will provide a critical test of the quark hadronization process. 
Further, their inclusion in the analysis of high precision experimental data  
will allow us to improve the description of the data and reduce the systematic 
errors in the extractions of the unpolarized and Collins fragmentation functions.
%%%%%%%%%%%%%%%%%%%%%%%%%%%%%%%%%%%%%%%%%%%%%%%%%%%%%
%%%%%%%%%%%%%%%%%%%%%%%%%%%%%%%%%%%%%%%%%%%%%%%%%%%%%
%%%%%%%%%%%%%%%%%%%%%   SECTION %%%%%%%%%%%%%%%%%%%%%%%%%%
%\section*{Acknowledgements}

{\it Acknowledgements.} -- This work was supported by the Australian Research Council through Grants 
FL0992247 
(AWT), CE110001004 (CoEPP) and by the University of Adelaide.

\vspace{-0.9cm}

\bibliographystyle{apsrev}
\bibliography{}

\end{document}